\documentclass{aastex61}
\usepackage{epstopdf}

\shortauthors{Willis \& Garrod}

\begin{document}
\raggedbottom
\title{Kinetic Monte Carlo simulations of the grain-surface back-diffusion effect}

\author{Eric R. Willis}
\affiliation{Department of Chemistry, University of Virginia, Charlottesville, VA 22904-4319}

\author{Robin T. Garrod}
\affiliation{Department of Chemistry, University of Virginia, Charlottesville, VA 22904-4319}
\affiliation{Department of Astronomy, University of Virginia, Charlottesville, VA 22904-4325}

\correspondingauthor{Eric R. Willis}
\email{ew2zb@virginia.edu}

\begin{abstract}
Rate-equation models are a widely-used and inexpensive tool for the simulation of interstellar chemistry under a range of physical conditions. However, their application to grain-surface chemical systems necessitates a number of simplifying assumptions, due to the requirement to treat only the total population of each species, using averaged rates, rather than treating each surface particle as an independent entity. While the outputs from rate-equation  models are strictly limited to such population information, the inputs -- in the form of the averaged rates that control the time-evolution of chemical populations -- can be guided by the results from more exact simulation methods. Here, we examine the effects of back-diffusion, wherein particles diffusing on a surface revisit binding sites on the lattice, slowing the total reaction rate. While this effect has been studied for two-particle systems, its influence at greater surface coverage of reactants has not been explored. Results from two Monte Carlo kinetics models (one a 2-D periodic lattice, the other the surface of a three dimensionally-realized grain) were used to develop a means to incorporate the grain-surface back-diffusion effect into rate-equation methods. The effects of grain size, grain morphology, and surface coverage on the magnitude of the back-diffusion effect were studied for the simple H+H reaction system. The results were fit with expressions that can be easily incorporated into astrochemical rate-equation models to reproduce accurately the effects of back-diffusion on grain-surface reaction rates. Back-diffusion reduces reaction rates by a maximum factor of around 5 for the canonical grain of $\sim$10$^6$ surface sites, but this falls to unity at close to full surface coverage.
\end{abstract}


\section{Introduction}

Chemistry occurring on the surfaces of interstellar dust grains is crucial to the formation of molecules in the interstellar medium \citep{HerbstvD}, including the most abundant interstellar molecule,  H$_{2}$ \citep{Gould}. The grain surface acts as a crucible for H$_{2}$ formation, allowing two adsorbed hydrogen atoms to meet by thermal diffusion and react, typically followed by desorption of the newly-formed H$_{2}$ molecule. Other grain-surface reactions occur similarly, with some involving activation energy barriers, although, unlike H$_2$, the products are generally expected to remain on the grain surface at low temperatures. 

Grain-surface chemistry was first incorporated into astrochemical kinetics models in the 1970s \citep{Pickles77a,Pickles77b}. Theoretical treatments of interstellar grain-surface chemistry have since become increasingly detailed (e.g. \cite{HHL}), including the formation of many more complex organic molecules. The most common method used to simulate the time-dependent evolution of interstellar chemistry is the so-called {\em rate-equation} (RE) method, whereby a system of ordinary differential equations (one for the abundance of each chemical species in the network) is solved using publicly-available solver routines. Construction of the differential equations requires the evaluation of average rates of reaction (and other processes); the calculated abundances thus correspond to average values, often interpreted as a time-average over a period in which the macroscopic conditions of the system remain constant. 

Rate-equation treatments are very accurate when applied to pure gas-phase chemistry, but can fail in certain regimes when the method is used to simulate grain-surface chemistry, due to the discrete nature of the grains. When the average population of reactive species on grain surfaces falls below order unity, stochastic effects can become important. This particular problem can be remedied reasonably well using modified rate equations, MRE \citep{Rob08}.

The RE and MRE methods provide a fast and efficient means to simulate coupled gas and grain chemistry, but the application specifically to grain-surface reactions still retains certain inaccuracies related to the use of average reaction rates, which are determined by the rates of diffusion of surface species. The typical approach \citep{HHL} is to calculate the absolute reaction rate between diffusive species, $i$ and $j$, using the expression:
\begin{equation}
R_{ij}=\kappa_{ij}k_{reac}N_{i}N_{j}
\end{equation}
where $\kappa_{ij}$ is a reaction efficiency related to the activation energy, $N_{i}$ and $N_{j}$ are the populations of species $i$ and $j$ on the grain surface, and $k_{reac}$ is defined as follows:
\begin{equation}
k_{reac}=\frac{k_{hop,i}+k_{hop,j}}{N_{s}}
\end{equation}
where $k_{hop,i}$ and $k_{hop,j}$ are the diffusion rates of species $i$ and $j$, respectively, from one site to an adjacent site, and $N_{s}$ is the number of binding sites on the surface. A similar expression is used in the case that $j$=$i$, such as that of the reaction H + H $\rightarrow$ H$_2$:
\begin{equation}
k_{reac}=\frac{k_{hop,i}}{N_{s}}
\end{equation}
The quantity $k_{hop}/N_{s}$ is sometimes referred to as the {\em scanning rate}, and is interpreted as the rate at which a diffusing particle succeeds in visiting all surface sites, which, in the case of only two particles on the grain surface, would be equal to the average number of hops required for two reactants to meet. However, this implicitly assumes that the particle visits each site only once. A more rigorous derivation for Eqs. (1) -- (3) considers the absolute reaction rate, $R_{ij}$, to be determined by the rate of hopping of species $i$ into an adjacent site, $k_{hop,i}$, multiplied by the probability that the new site contains a reaction partner $j$. The latter probability is given by $N_{i,j}/N_{s}$, the fractional coverage of the surface by species $j$. A similar term may be constructed for diffusion by species $j$ into a site containing reaction partner $i$. 

By basing the overall reaction rate on the individual rate associated with a single hop, rather than the full chain of diffusion events leading to reaction, the above treatment ignores the possibility that the random walk of the diffusing species may include hops that return it to previously-visited binding sites. This effect, known as {\em back diffusion}, will act to slow down grain-surface reactions compared to the standard treatment used in astrochemical models.

There has been extensive research conducted on the theory of random walks in statistical physics and other fields, of which we cite a small fraction; \cite{Hatlee} investigated the problem of random walks on finite lattices using a Monte Carlo approach, in an attempt to investigate the effect of boundaries on processes related to chemical dynamics.  \cite{Botar} used Monte Carlo models to study reaction kinetics at different solute concentrations, while \cite{Allen} used a combination of analytical and Monte Carlo techniques to study the relationship between the diffusion coefficient and rate constant on square lattices. More recently, \cite{Paster} investigated a similar problem to that presented in this work, investigating the effect of stochastic initial conditions on the diffusion-reaction equation on grids in 1-3 dimensions.

The problem of random walks on interstellar dust grains in particular has been approached by various authors. \cite{Charnley} investigated the diffusion on grain surfaces in order to challenge the so-called `two-coreactant restriction.' Charnley used results already known for random walks, showing that for a surface with 10$^{6}$ binding sites, reaction would be slowed by a factor of $\sim$4.9 when including back-diffusion, indicating that the scanning rate should be adjusted accordingly. This derivation considered a single particle diffusing over the grain, so surface-coverage effects were not included.

\cite{Krug06} also investigated a modification to the scanning rate in models of grain-surface chemistry. They considered a spherical grain with two surface particles adsorbed. One particle was forced to be stationary, while the other was allowed to diffuse, but in the case where all surface sites are identical, this arrangement is equivalent to having two mobile (identical) particles. They then calculated the exact scanning rate for this situation, defined in terms of an encounter probability for the two particles (see also \cite{LohmarKrug09}). This encounter probability was found to depend upon both the diffusion and desorption rates of the mobile atom. The new scanning rate was broken into two regimes of interest: small grains and large grains. The true scanning rate was found to be reduced in comparison to the conventional approximation, with this reduction being greatest for large grains, but still significant on smaller grains. Overall, the factor by which this exact scanning rate slowed down reactions was between $\sim$2-5, depending on grain size. The follow-up paper by \cite{Krug09} presented a more approximate solution to the problem for ease of inclusion into astrochemical models, conducting kinetic Monte Carlo simulations that showed that the approximation accurately reproduced the behavior seen in the exact solution. Despite the promise of this work, it has not yet been tested in an astrochemical model. However, the derivation of the encounter probability undertaken by \cite{Krug06} considers just two particles on the grain surface, while interstellar dust particles can accumulate larger numbers of reactive atoms and molecules.

In this paper, we explore the dependence of the back-diffusion effect not only on the number of surface binding sites on the grain, but on the surface coverage of reactants, using kinetic Monte Carlo methods that can simulate a range of astrophysically-relevant conditions and surfaces. We use both a 2-D model of a square surface with periodic boundary conditions, and a fully three-dimensional model of a grain, in which the surface binding sites and the associated directions of diffusion are determined by the arrangement of the atoms that make up the grain surface (see Garrod 2013). Such methods are useful in solving this problem, as they allow the positions of particles on a grain surface to be traced explicitly, allowing averaged path lengths (in terms of the number of hops) to be calculated over large numbers of diffusion/reaction events. The resultant averaged scanning rates may then be incorporated directly into standard rate-equation models. The use of numerical simulations such as these also lay the groundwork for their use in characterizing the kinetic properties of rough, amorphous surfaces.

In the simulations and analysis presented here, we define the back-diffusion factor, $\phi$, as the factor by which reactions are slowed by the back-diffusion effect, as compared with the standard rate-equation formulation of Eqs. (1) -- (3). 
With this definition in mind, it is also convenient to define an effective number of surface sites on the grain, $N_{Eff}$, which may, as with Eq. (3) in the case of two particles on the surface, be identified with the average number of hops required for reaction to occur. These quantities are related thus:
\begin{equation}
{N_{Eff}} = \phi {N_{s}}
\end{equation}
where $N_{s}$ is the actual number of binding sites on the surface. We present fits to the computational calculations of average rates, in order to provide practical determinations for $\phi$ that may be easily employed in astrochemical models.

In section 2, we describe the computational methods. In section 3, we present results. Section 4 contains a discussion of the results, and section 5 presents the conclusions of this study.

\section{Methods}
\subsection{Simple flat-surface model}
The surface chemistry simulations presented in this paper were performed using two kinetic Monte Carlo models. The simpler model uses a flat surface with rectangular lattice geometry in a square grid. This surface has a user-determined size with periodic boundary conditions. In this model, the number of binding sites, and their location, are directly specified. A user-selected number of target atoms is then deposited onto the surface in randomly-chosen binding sites. These target atoms are not allowed to diffuse, and are analogous to heavier grain-surface atoms that diffuse very slowly relative to hydrogen, such as oxygen. The last particle deposited onto the surface is mobile, analogous to a reactive hydrogen atom. This particle diffuses via a random walk. All four diffusion directions have equal probability in the simulation, and are chosen randomly. When a reaction occurs, the number of hops to achieve reaction is recorded, and the surface is cleared of particles. The model does not trace the time, only the number of hops, and as such is insensitive to the value of the diffusion barrier. The process then restarts, until a user-specified number of reactions is recorded. In this way, it is possible to determine the effect of back-diffusion as a function of surface coverage for varying surface sizes. Results were obtained for three sizes (10,000, 1,000,000, and 4,000,000 sites), sampling a wide range of surface coverages by altering the number of deposited stationary particles.

\subsection{MIMICK}
The two-dimensional model has shortcomings; firstly that only one particle is allowed to diffuse on the surface, and secondly that the surface is flat and periodic, rather than being the bounding surface of a three-dimensional dust grain. In order to overcome both these deficiencies at once, the off-lattice Monte Carlo kinetics model MIMICK (Model for Interstellar Monte-Carlo Ice Chemical Kinetics) was used (\cite{Rob13}). MIMICK allows for the simulation of chemistry on grains of user-defined size and morphology. The positions of all particles are determined explicitly based upon the local potential minima on the surface, so there are no pre-defined lattice sites, thus making it a true off-lattice model. In this paper, the chemical network has been vastly simplified, with atomic hydrogen being the only species allowed to accrete onto the grain surface from the gas phase. This model explicitly traces the passage of time, and uses an explicit diffusion barrier. {However, while exact pairwise potentials are used to determine the positions of the surface potential minima, all diffusion barriers are set to $\sim$510.6 K (the value for amorphous carbon determined by \cite{Katz}), to replicate the conditions assumed by \cite{Krug09}.}

\subsubsection{Grains}
Two grain morphologies were used with MIMICK to investigate the effect that grain-surface morphology has on the back-diffusion factor. The first was a cubic grain, created using a simple cubic lattice. Each binding site (i.e. potential well) on the grain has four diffusion paths leading out of it, which is identical to the flat surface used in the simple two-dimensional model. It was not possible to create a simple flat surface for use in MIMICK, due to the difficulty of incorporating periodic boundary conditions into this more complex model. Thus the cube was chosen as a reliable comparison to the flat surface, to determine if the transition from the flat surface to a confined grain geometry had an effect on the back-diffusion factor.

The second grain used in this study assumed the shape of a bucky-ball, created with the aid of the DOME program.\footnote{www.antiprism.com/other/dome} Coordinates for each bucky-ball were generated using DOME, and these coordinates were then transformed into input compatible with MIMICK. Some error may be introduced as a result of the coordinate transformation, causing the spacing between atoms to vary on small scales throughout the grain. However, this does not impact the results presented here, due to the adoption of a fixed diffusion barrier for all sites and directions. The bucky-ball was chosen because it is spherical, with all binding sites being essentially equivalent, and as such is analogous to the spherical grain geometry assumed by rate-equation codes. Each binding site on the bucky-ball grain has a hexagonal geometry. An image of both grains is shown in Figure 1, created using POV-Ray.\footnote{www.povray.org} 

\begin{figure}[ht!]
\plotone{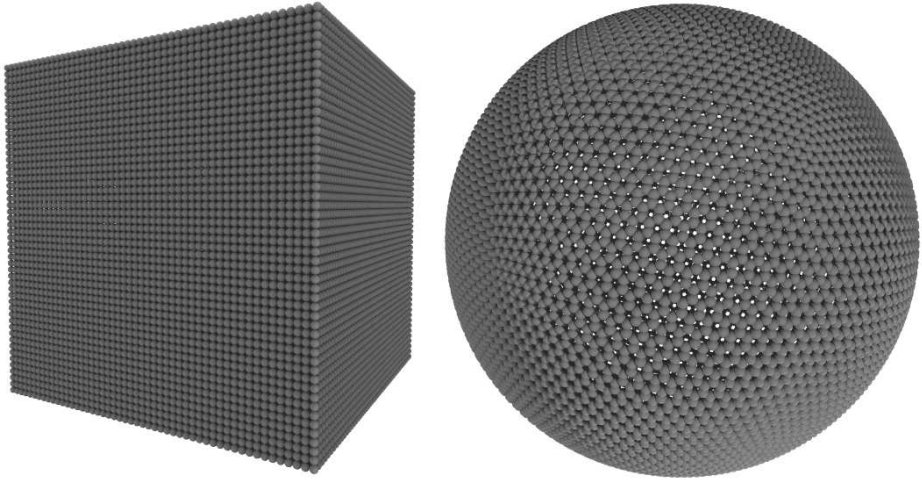}
\caption{The two grain morphologies used in this study. The grain on the left is the cubic grain with 15,000 sites (side length of 40 \AA), while the grain on the right is the bucky-ball grain with 6,757 sites (radius of 95 \AA). The grains are not to scale with each other.}
\end{figure}

For the cube morphology, grains of 15,000, 135,000 and 301,056 surface binding sites were created. For the bucky-ball morphology, grains of 6,757, 32,710 and 300,665 surface binding sites were used. Table 1 contains a short summary of all surfaces used in this study. The number of binding sites on each grain was determined computationally, by allowing a test particle to sample all positions on the surface. All MIMICK simulations were undertaken with a grain temperature of 18 K and a gas temperature of 100 K, following \cite{Krug09}

\begin{deluxetable}{ccccc}
\tablecaption{Summary of surfaces used in this study.}
\tablecolumns{5}
\tablewidth{0pt}
\tablehead{
\colhead{Morphology} & \colhead{$N_{S}$} & \colhead{Radius/side length (\AA)} & \colhead{Nearest-neighbor distance (\AA)} & \colhead{Code}}
\startdata
Flat surface & 10,000, 1,000,000, 4,000,000 & N/A & N/A & Flat-surface model \\
Cubic & 15,000, 135,000, 301,056 & 40, 120, 179.2 & 3.2 & MIMICK \\
Bucky-ball & 6,757, 32,710, 300,665 &  95, 215, 650 & $\sim$4-5 & MIMICK \\
\enddata
\end{deluxetable} 

\subsubsection{MIMICK, one mobile particle}
A vastly simplified version of MIMICK was first used in order to test the effect of using a confined three-dimensional grain geometry instead of a flat surface. In this model, a specified number of particles was accreted onto the grain surface. The last particle was allowed to diffuse, while the others were held static, following the conditions of the flat, two-dimensional model. The number of hops by the lone mobile particle was counted before reaction occurred. All particles were then removed from the grain surface. This process was repeated a sufficient number of times to account for fluctuations in particle position, and an average number of hops was taken. This was done for the cubic grain with 15,000 binding sites, and the bucky-ball grain with 32,710 binding sites.

\subsubsection{MIMICK, all particles diffusing}
To investigate the behavior of a system in which all surface reactants are allowed to move, simulations were run using the standard MIMICK code with a few small modifications. Firstly, desorption from the grain surface was prohibited, in order to isolate back-diffusion from the effects of competing processes. This is one point of deviation between our study and that of \cite{Krug09}. However, desorption is not expected to have a significant effect on this hydrogen-only chemical network, if realistic desorption barriers are chosen. In the case of a grain at 10K, assuming $E_{b}/E_{D}$=0.35 and E$_{D}$=450K (Garrod \& Pauly 2011), a particle is expected to undergo on average $\sim$5 x10$^{12}$ hops before thermal desorption, and $\sim$10$^{20}$ hops before photo-desorption (assuming A$_{V}$=10). Both figures are much higher than the average number of hops a particle performs in our simulations. It is possible that thermal desorption may become a competing process at higher grain temperatures (e.g. at 18K, a particle will undergo an average of $\sim$10$^{7}$ hops before thermal desorption). In addition, photodesorption may be important at the edge of a cloud, where A$_{V}$ values are lower (although the the precise rates for the photodesorption of atomic species is not well-constrained). However, in standard dense cloud conditions, these processes are not expected to compete with diffusion, and as a result are not included.

In a further simplification, Eley-Rideal reactions, in which an accreting particle lands directly on top of an adsorbed particle and immediately reacts, have been removed from this version of MIMICK; accretion events that would otherwise lead to immediate reaction are repeated until a non-reactive accretion is achieved.  Although Eley-Rideal reactions are not important at low surface coverages, they can become influential at higher coverages, thus affecting reaction rates in ways that are not tied to back-diffusion. 

Thirdly, the reaction process has been altered, so as to reproduce the behavior assumed in the rate-equation treatment, whereby reaction occurs when a reactive particle hops into the same binding site as a reaction partner. The standard version of MIMICK allows particles to react when they are within sufficient distance to interact (often in adjacent binding sites), rather than being required to ``share'' the same potential well.

In these models, unlike those with only one mobile particle, the accretion and diffusion processes are each allowed to occur continuously, such that there are natural fluctuations in the instantaneous populations of reactants on the surface. Data are recorded only once the mean population becomes stable.

\section{Results}
\subsection{Flat-surface model}

The flat-surface model was used first, in order to investigate how back-diffusion varies with surface size and coverage for a single diffusive particle. We ran models for each of the three square lattices (10,000, 1,000,000, and 4,000,000 binding sites). Models were run multiple times for the same set of conditions (using different random number seeds) until a stable mean number of hops was achieved. 

The results of these simulations are shown in Figure 2, in which the back-diffusion factor, $\phi$, is plotted against $N_{s}/N_{B}$, the inverse of the surface coverage, where $N_{B}$ is the number of non-diffusive reactive particles. Each value of $\phi$ is determined by dividing the average number of hops per reaction by $N_{s}/N_{B}$.

\begin{figure}[h!]
\plotone{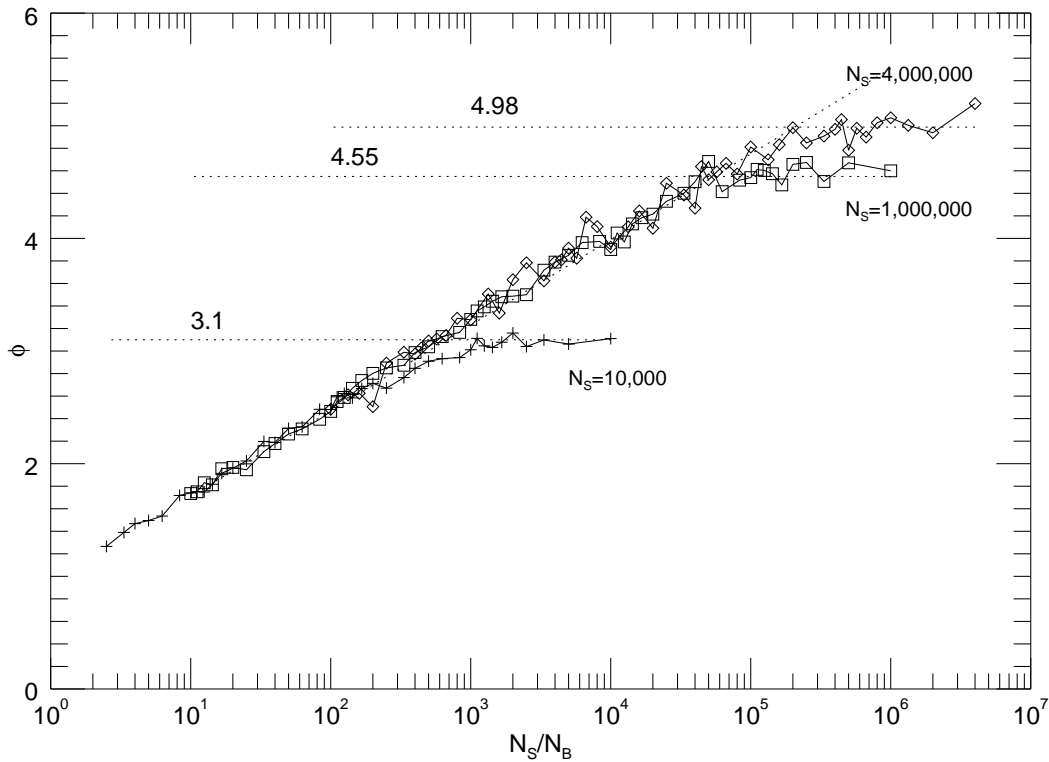}
\caption{Data from the flat-surface models, showing the back-diffusion factor versus the inverse surface coverage. Data from three different surface sizes are plotted: 10,000 sites (crosses), 1,000,000 sites (squares) and 4,000,000 sites (diamonds). The values of the plateaus for each size are also shown.}
\end{figure}

The results show a clear correlation between the inverse surface coverage and the strength of the back-diffusion effect. As the number of particles on the surface decreases in the figure (from left to right), the effect of back-diffusion increases. All three surface sizes that were tested, each plotted as a separate curve, follow the same relationship until a certain threshold of low-coverage is reached, beyond which the value of $\phi$ hits a plateau; a different plateau value is obtained for each surface. Conversely, as the coverage on the surface increases (moving leftward in Fig. 2), back-diffusion becomes less important until it ultimately loses all influence at an inverse surface coverage of 1, at which point reaction requires only a single hop. This behavior is seen for all surface sizes. Thus, the size of the surface is only important (independently of surface coverage) at the extreme low-coverage end of the data.

We fit these data, such that the main portion of the curve for each grain size is described by the expression:
\begin{equation}
\phi = 0.95 + 0.315 \ln(\frac{N_{s}}{N_{B}})
\end{equation}
Plateau values are dependent on the grain size, and are fitted by the expression:
\begin{equation}
\phi = 0.2 + 0.315 \ln(N_{s})
\end{equation}

\subsection{MIMICK results, one mobile particle}
To determine whether the transition from a flat surface with periodic boundary conditions to a confined grain geometry has a significant effect on the back-diffusion factor, we use the MIMICK model with only one reactant allowed to move, as described in \S 2.2.2. The results are shown in Fig. 3, overlaid with the flat-surface data from \S 3.1. The grains used here are the cubic grain with 15,000 binding sites, and the bucky-ball grain with 32,710 binding sites.

\begin{figure}[h!]
\plotone{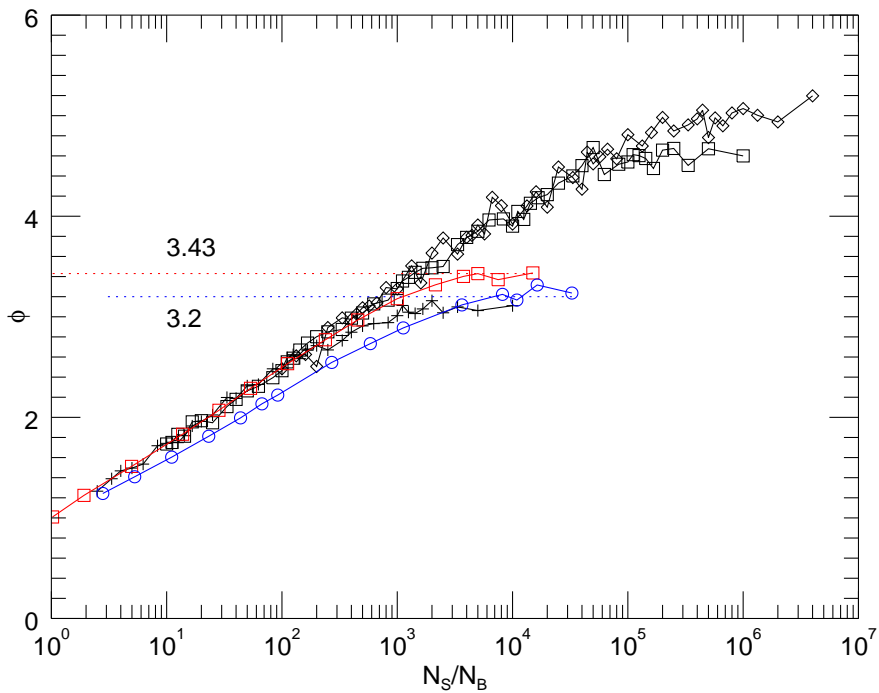}
\caption{Data from MIMICK with one mobile particle overlaid with flat-surface data (shown in black); cube-grain data are shown as red squares, bucky-ball data as blue circles. Plateau values for each grain are marked. The cubic grain used here has 15,000 surface binding sites, while the bucky-ball grain has 32,710.}
\end{figure}

The cube results are very similar to those from the flat-surface model, showing the same relationship between back-diffusion and surface coverage on the main part of the curve. This indicates that the change from periodic boundary conditions to a surface on a three-dimensional grain has no effect under most conditions, for identical binding site geometries. The exception to this, however, is the plateau value at low coverage, which is observed to be somewhat higher ($\sim$3.43) than the calculated value for a flat surface of the same size, using Eq. 4 ($\sim$3.23). This is likely a consequence of the different diffusion paths that become available to the mobile particle when the transition is made from periodic boundary conditions to a confined grain geometry. In the latter case, the shortest path of diffusion around the lattice (the straight-line path) becomes longer, which increases the number of hops that it takes for reactive particles to find each other in the low-coverage regime. For the example of the cubic grain (15,000 sites), the fewest hops required to move around the surface in one direction and return to the starting point would be 200 hops, while a flat surface of the same number of sites would have a shortest diffusion path of $\sim$122 hops.

The bucky-ball results suggest that grain and binding-site geometry also impacts the back-diffusion effect. The slope of the curve is shallower and, in spite of the fact that this particular grain has more than twice as many binding sites as the cube, the back-diffusion factor reaches a maximum at a lower value of $\sim$3.3. This is believed to be a result of the different binding site geometries of the two grains. The cubic grain has 4 nearest neighbors, while the bucky-ball has 6.

\subsection{MIMICK results, all particles mobile}
The cube and bucky-ball morphologies were investigated using the full version of MIMICK, with all particles allowed to diffuse, as described in \S 2.2.3.
The results for the cubic and bucky-ball grains are presented in Figures 4 and 5, respectively, which again plot the back-diffusion factor, $\phi$, against inverse surface coverage. Because all particles are allowed to move, the back-diffusion factor cannot be determined simply by counting the number of hops. Instead, $\phi$ is now calculated directly, by determing the overall rate of reaction (reactions per unit time) produced in the Monte Carlo simulations, $R_{MC}$, and dividing by the expected rate-equation rate, $R_{RE}$, calculated using the time-averaged population produced by the Monte Carlo models. By running the models with different values of the accretion rate of H onto the grains, a wide range of grain-surface coverages are tested.

\begin{figure}[h!]
\plotone{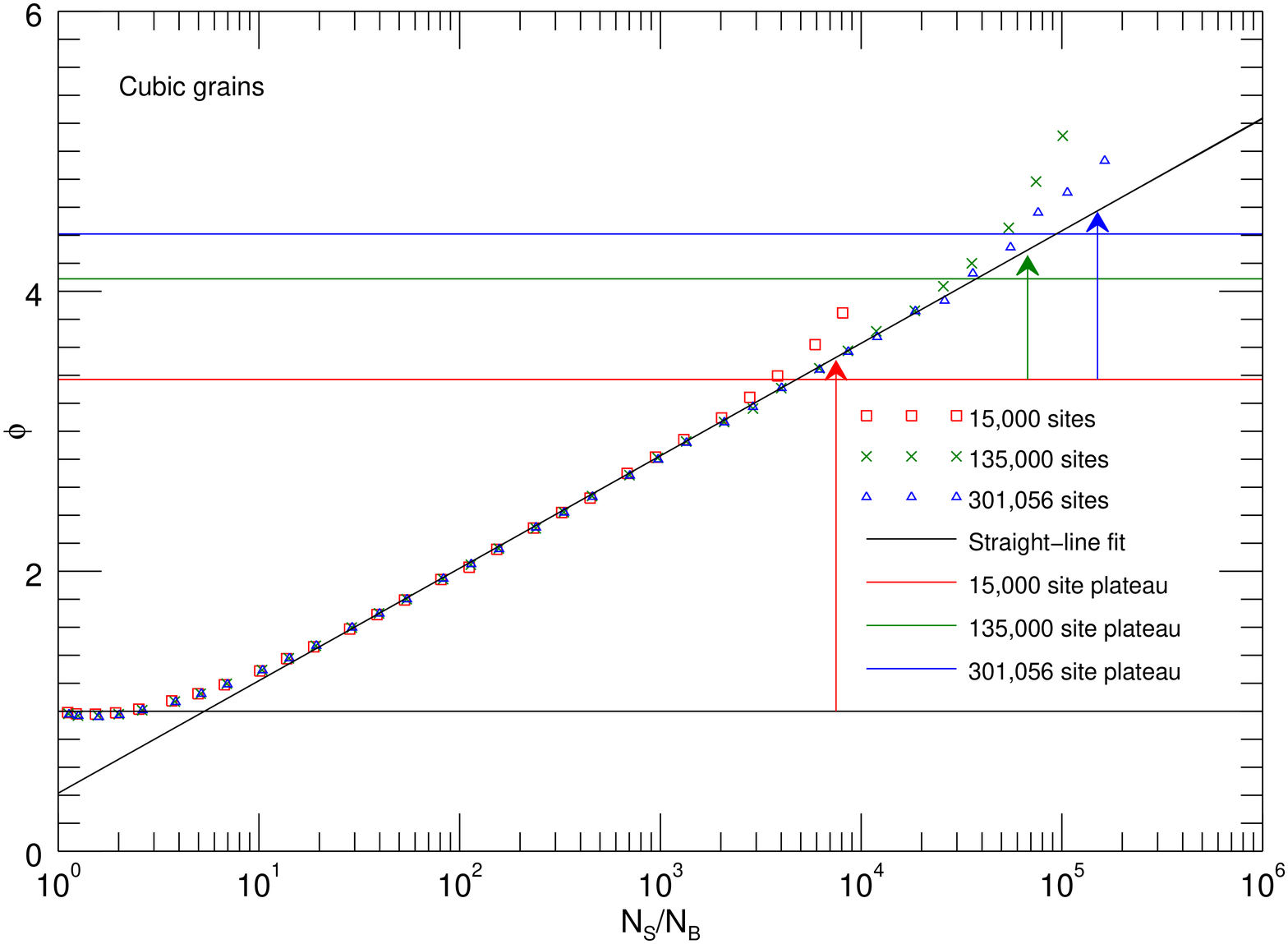}
\caption{Results for cubic grains, obtained using the full reaction treatment in MIMICK, allowing all particles to diffuse. The arrows are color-coded to correspond to the data points for the same grain size, and indicate the points on the line corresponding to an average of two particles on each grain.}
\end{figure}

\begin{figure}[h!]
\plotone{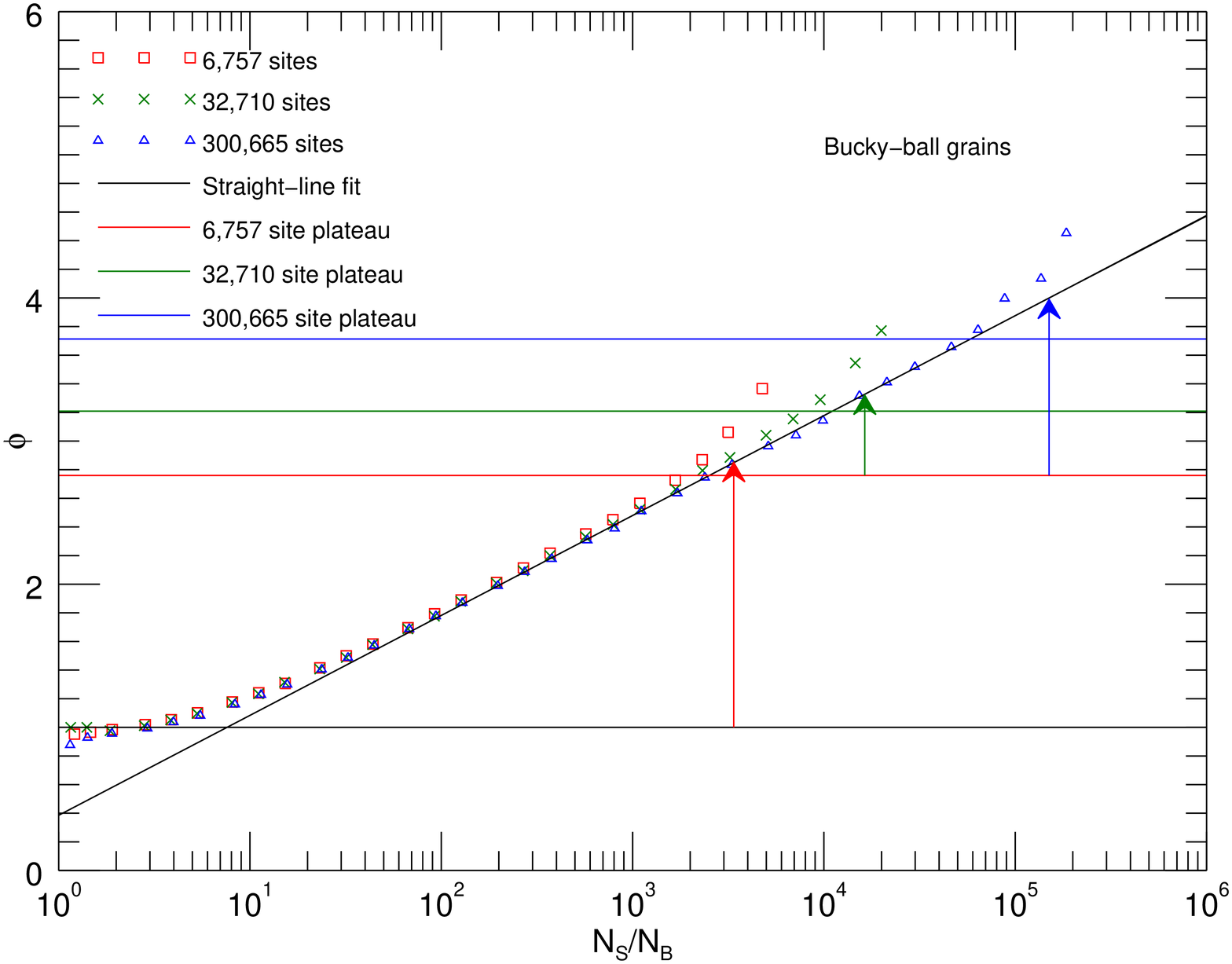}
\caption{Results for bucky-ball grains, as per Fig. 4.}
\end{figure}

The curves for all grain sizes and morphologies shown in Figs. 4 \& 5 achieve a back-diffusion factor of 1 at a value $N_{s}/N_{B}$=1; however, this value is reached at lower coverages, and the central portions of the curves no longer point to $\phi=1$, unlike the results of \S 3.1-3.2. The intermediate-coverage behavior for both morphologies can be fit using a logarithmic function dependent on the surface coverage, as before, and the main part of the curve obeys the same expression for all grain sizes within either type of grain morphology (see Table 2.) The absolute values of the back-diffusion factor are lower here than those observed for either the flat-surface model or the MIMICK models with one mobile particle. Again, the bucky-ball produces lower values of $\phi$ than the cubic grain.

\subsection{MIMICK, two-particle models}
In Figures 4 \& 5, as the inverse coverage reaches very high values, the results diverge from logarithmic behavior at a coverage on the order of five particles, for each grain size considered. Values of $\phi$ become yet greater as the average number of particles per grain decreases below this value, and the plateaus seen previously are not apparent in any of the results. 

This divergence from the behavior seen in the other models is actually an artifact of the method used {\em post hoc} to determine the total rate of reaction produced by the Monte Carlo models, $R_{MC}$, which is used to calculate $\phi$. Because $R_{MC}$ is evaluated by dividing the total number of reactions by the amount of time passed, this means that, for very low coverages, much of this time corresponds to occasions when the population of reactants on the grain is either 0 or 1, meaning that no reaction can occur. This ``dead-time'' would correspond to many hopping events that do not contribute to the reaction process, and should therefore not be included in the determination of the back-diffusion effect. Note that low coverages in these models correspond to low {\em mean} populations, which may be less than two, while reactions can only occur when the instantaneous population state is greater than or equal to 2.

A separate model was used in order to study the behavior on the grains when only 2 particles were present. In these two-particle simulations, accretion was always halted as soon as two particles were present on the surface, and was allowed to resume following each reaction. The number of hops between reactions was counted, thus eliminating any time dependence. These two-particle values may be considered as cutoffs, or plateau values, for the back-diffusion factor for each grain size and type. Straight lines at these values are overlaid on the data in Figs. 4 and 5, color-coded for each data set. If a calculated back-diffusion factor is found to be above these values, it should be corrected down to the two-particle value for that grain size. Generalized fits are again presented in Table 2, for both the main part of the curves and the plateaus.

\begin{deluxetable}{ccc}
\tablecaption{Fits for each grain morphology.}
\tablecolumns{3}
\tablewidth{0pt}
\tablehead{
\colhead{Morphology} & \colhead{Straight-line fit} & \colhead{Plateau fit}}
\startdata
Flat (single mobile particle) & 0.315 $\ln{\frac{N_{S}}{N_{B}}}$ + 0.95 & 0.315 $\ln{N_{S}}$ + 0.2 \\
Cubic                                  & 0.3489 $\ln{\frac{N_{S}}{N_{B}}}$ + 0.4146 & 0.3423 $\ln{N_{S}}$ + 0.07224 \\
Bucky-ball                           & 0.3032 $\ln{\frac{N_{S}}{N_{B}}}$ + 0.3856 & 0.2496 $\ln{N_{S}}$ + 0.5795 \\
\enddata
\end{deluxetable}

\subsection{Comparison of fits to rate-equation results}

The derived fits, shown in Table 2, are simple to incorporate into rate-equation models. The back-diffusion factor should be calculated on-the-fly, as the grain-surface coverage changes during a simulation, while the plateau value for the grain size in question need only be calculated once. If the back-diffusion factor is greater than the plateau value, or less than 1, it should be corrected to those corresponding values. Finally, the resulting back-diffusion factor should be incorporated into the reaction rate. This is done by modifying Equation 2, introducing the back-diffusion factor into the formula as follows:
\begin{equation}
k_{reac}=\frac{k_{hop,i}+k_{hop,j}}{\phi N_{s}}
\end{equation}
where $\phi$ is the back-diffusion factor, with a similar correction made to Eq. 3.

To demonstrate the implementation of the back-diffusion correction into astrochemical models, we carry out simulations of H$_2$ formation on the 15,000-site cubic grain using both the MIMICK Monte Carlo model, and a rate-equation model that incorporates the rate corrections shown in Table 2. In both models, we calculate the steady-state value of the mean population of H atoms on the grain, for each value of a set of accretion rates. Because we disregard desorption of H in these models, the rate of H$_2$ formation is necessarily half the rate of H accretion in every case. Modified rates are also included in the rate-equation model, following \cite{Rob08}, such that at low grain-surface populations, the following equation is used for the H+H reaction rate:

\begin{equation}
R_{mod}=R_{acc}N_{B}
\end{equation}

\begin{figure}[h!]
\epsscale{1}
\plotone{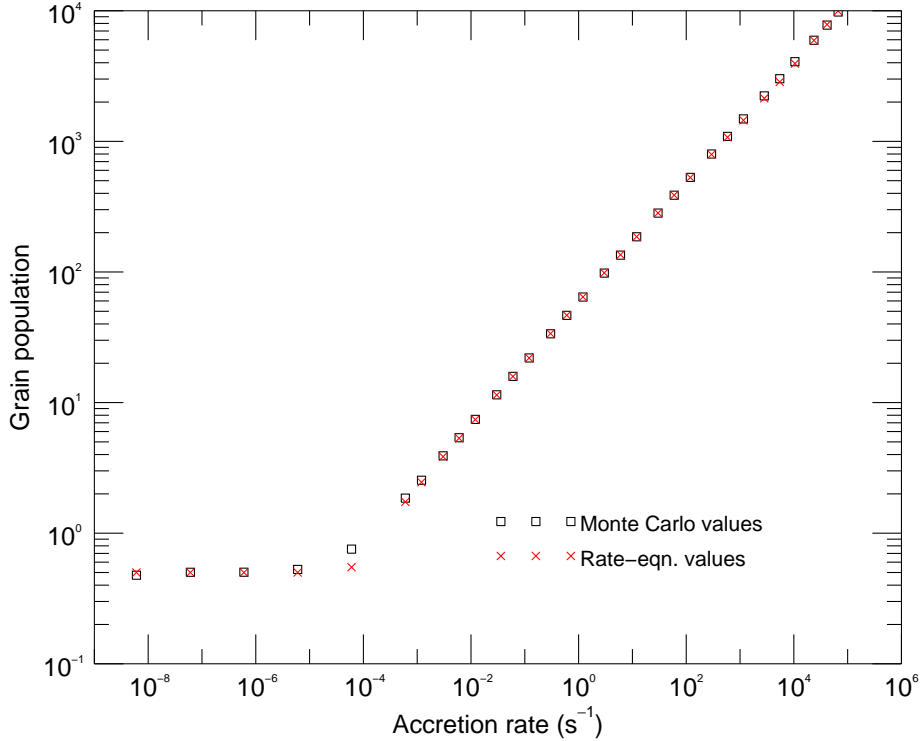}
\caption{Comparison of grain-surface populations obtained from MIMICK and those obtained from a simplified rate-equation treatment utilizing the back-diffusion correction presented in this paper. This comparison is done using the cubic grain with 15,000 sites.}
\end{figure}

Figure 6 shows the results from each of the two models. The back-diffusion correction imported into rate-equation treatment shows very good agreement at almost every point, with the exception of the fifth point from the left. Here, the Monte Carlo model has an average grain-surface population of 0.7549, which puts it within the modified-rate regime, but close to the threshold between rate-equation and modified-rate behavior used by \cite{Rob08}. Its proximity to this threshold appears to be the source of the disagreement between the two methods, rather than the back-diffusion correction. At all other points, disagreement between the two methods rarely exceeds 5\%, with the majority of points being within 1\% of each other. Similar levels of agreement are seen for all grains tested in this paper, showing that this method is valid and will produce the appropriate results in real astrochemical models. Note that the \textit{production rates}, which are not shown, are an exact match between models in all cases, due to the imposed steady-state condition.
\\ \\

\section{Discussion}

It is instructive to compare the low-coverage results obtained here with the back-diffusion factors determined by \cite{Krug09} for the two-particle case. The plateau values displayed in Figs. 4 \& 5 are seen to be in reasonable agreement with the results of Krug et al. (as seen in Fig. 1 of that paper, specifically their values with R$_{des}$:R$_{diff}$ of 10$^{-6}$). For example, the cubic grain with 15,000 sites shown in Fig. 4 has a plateau value of $\phi$$\simeq$3.37, while a spherical grain with square-lattice geometry and a similar number of sites is found by \cite{Krug09} to have a back-diffusion factor of $\sim$3.33, suggesting that both treatments are similar in the low-coverage regime, in spite of the morphological differences between the grains in either model. Using the fit (Table 2) to the results for our spherical grain with hexagonal-lattice structure (Fig. 5), and assuming 15,000 sites, a back-diffusion factor of $\sim$2.98 is obtained. Clearly, at low coverage, it is the local surface structure (i.e. number of diffusion pathways available) that is the primary influence on back-diffusion, rather than the choice of overall grain morphology (cube vs. sphere). This effect has been previously studied and observed to have a similar impact on hydrogen recombination efficiency by \cite{Chang}. In that paper, the authors investigated hydrogen recombination on lattices with rectangular and hexagonal binding site geometry using a continuous-time random-walk model. They found that, for homogeneous surfaces like those we are interested in, the hexagonal lattice shows a greater hydrogen recombination efficiency. This indicates a lesser impact from back-diffusion, similar to our results.

It may also be seen that the value of $\phi$ at the plateau (i.e. very low coverage) is exclusively determined by the size of the grain, assuming the same surface structure and morphology. However, the transition from a flat, periodic surface to a surface that bounds a three-dimensional grain produces a somewhat larger degree of back-diffusion. This transition does not, however, appear to have any significant effect other than at the plateau. The differences seen at all coverages between the flat-surface treatment and the bucky-ball grain appear to be entirely caused by the alternative local structure, as do the differences between the bucky-ball and the cubic grain. It is possible that a more extreme grain morphology could show a greater deviation; for example, one may consider a hypothetical grain composed of two contiguous spheres, with just a small number of binding sites connecting the two parts. Such would be an interesting case for future study.

At the higher coverages that are examined only in our models, in the case where just one surface particle is mobile (Fig. 3), it is also apparent that the plateau value of $\phi$ (i.e. the extreme two-particle value) is no longer applicable beyond some threshold, and that the back-diffusion factor then begins to fall with increasing numbers of particles on the surface, reaching a value of 1 at full surface coverage. A similar trend is seen for the case where all particles are mobile, but in this case $\phi$ reaches unity (i.e. no back-diffusion effect) at a surface coverage less than one (i.e. $N_{s}/N_{B}>1$). This effect may be understood by considering that at high coverages, there is a very large probability that multiple particles may be within just a single hop of a reaction partner at any particular moment. In the one-mobile-particle case, the back-diffusion effect manifests itself entirely through the diffusion path of a single particle, which, when averaged over many separate reaction simulations, requires that the diffusing particle be entirely surrounded by reaction partners to ensure that no back-diffusion occurs. One may surmise that in the case where all particles are mobile, back diffusion should become neglible when, on average, the inverse surface coverage is approximately equal to the number of diffusion pathways out of a binding site. Indeed, the main part of the curve in Figs. 4 \& 5 may be extrapolated to reach $\phi=1$ at values of $N_{s}/N_{B}$=5.4 and 7.6, respectively, for the cubic case (4 diffusion paths) and the bucky-ball case (6 paths). The number of paths out of a site also affects the gradient of the main part of the curve, but the difference is relatively modest, with more pathways resulting in a slightly reduced back-diffusion effect.

Note that the incorporation of the back-diffusion factor into the rate-equation method is only necessary for standard rate equations in the simple H$_{2}$-formation system explored here. When modified rates become operative, diffusion is no longer part of the rate calculation, and so back-diffusion becomes unimportant. In these instances, accretion is the defining timescale for reaction rates. In a full modified-rate treatment that considers desorption or other processes that may compete with reaction, the back-diffusion term must be included in the relevant competition terms, as per \cite{Rob08} (Eqs. 17-22 in that paper). This is achieved simply by applying the back-diffusion factor to the number of hops required for reaction, as per Eq. (7).

Desorption and Eley-Rideal processes were intentionally removed from the treatments presented in this paper, in order to isolate the explicit effects of back diffusion. Desorption would have an effect on back-diffusion in regimes in which the desorption barrier is not much higher than the diffusion barrier. This is described in the results of \cite{Krug06}, where the desorption barrier is $\sim$1.3 times the diffusion barrier. However, more commonly-chosen values place the desorption barrier 2-3 times greater than the diffusion barrier \citep{GarrodPauly}, thus its influence on back-diffusion would be minimal, as the diffusion rate would dominate the desorption rate. The inclusion of Eley-Rideal reactions would have no significant impact on back-diffusion, as reaction via this process only becomes important at very high coverages. Under these conditions, back diffusion is already minimal.

In a broader astrochemical context, the maximum back-diffusion factors achieved in the simulations presented here (in the range of 3-5), will have a significant scaling effect on grain-surface reaction rates. This is particularly true when systems not in the accretion limit are considered. In such a regime, the rates of reaction on the grain surface are determined by the diffusion rates, and thus scaling the diffusion rates down will have an appreciable effect on reaction rates. In the accretion limit, where grain-surface reaction rates are determined by accretion rates, this back-diffusion factor should not have a strong effect on the simple system presented here. Its importance in more complicated chemical networks in the accretion limit remains to be studied.

\section{Conclusions}
Two Monte Carlo models have been used to investigate the effect of back-diffusion on interstellar grain surfaces. Three surface morphologies have been used: a flat, periodic surface, a cubic grain, and a bucky-ball grain. The effect of back-diffusion has been studied as a function of grain-surface coverage, grain size and lattice morphology, and has been shown to slow down surface reactions by as much as a factor of 3-5, depending on these parameters, in line with past work. Crucially, it is found that the degree of back diffusion is strongly dependent on the surface coverage, while at low coverages a maximum back-diffusion factor is reached, whose value is dependent solely on the size of the grain (i.e. number of binding sites).

The main practical results of this paper are given by the fits provided in Table 2. These allow the back-diffusion factor to be incorporated easily into astrochemical models. While the fits concern several grain morphologies, it is likely that real interstellar grains  will show more complicated structures on multiple size scales. Amorphous surface structures on microscopic scales would result in a back-diffusion behavior that is more complicated than what is presented here, and which would depend on the strength of individual binding sites or classes of site. However, the fits that we provide here represent a marked improvement in accurately describing grain-surface chemistry over all surface coverages. Traditional rate-equation models of astrochemistry assume spherical grains, and as such it is expected that the bucky-ball data will be most useful for this purpose. However, the results from the cubic grain are not greatly divergent and would be quite adequate to match a surface that is better described by the square-lattice structure at a local level, as it is this structure that generally has the most influence on the back-diffusion factor. Investigations of more complex grain morphologies, as well as more complex chemical networks, are left for future studies.

There are 3 main features to the back-diffusion fits:
\begin{enumerate}
\item At high coverage, the back-diffusion factor reaches a value of 1, indicating that back-diffusion is not important in this regime, due to the high probability of a reactant meeting a reaction partner within a few (or one) hops.
\item The majority of the data, between the high-coverage and low-coverage regimes, can be described by a logarithmic function, dependent upon the surface coverage.
\item The back-diffusion factor reaches a maximum at a population of 2 particles on the grain surface, and these plateau values can also be described using a logarithmic function, dependent upon the surface size.
\end{enumerate}

This work builds upon the results of \cite{Krug06} and \cite{Krug09}. The plateau values for back-diffusion correspond reasonably well with the values found by those authors. However, deviation from these values is seen in the majority of our data, showing that accounting for grain-surface coverage is also important. This is of particular use when incorporated into astrochemical models, where a wide range of grain-surface coverages are encountered.

A further area of research on this topic relates to the usual assumption that a reaction occurs on the surface when the two reactants meet in the same binding site. In fact, in the physisorption regime in which these diffusive models are valid, each binding site represents a minimum in the potential that describes the surface. That potential will be affected by the presence of a reactant, such that possible reaction partners in adjacent binding sites on a regular lattice would already share a potential well, allowing them to immediately react (as assumed by Garrod 2013). Preliminary models using our Monte Carlo treatment suggest that the removal of the requirement for this final step in the diffusion path may have a significant effect on the back-diffusion effect.

\acknowledgements
The authors would like to thank the anonymous referee for helpful comments. The authors would also like to thank Eric Herbst for useful discussions. RTG thanks the NASA Laboratory Astrophysics program for funding through Grant NNX15AG07G. 

\software{MIMICK (Garrod 2013), DOME (www.antiprism.com/other/dome), POV-Ray (www.povray.org)}

\end{document}